\documentclass[%
 reprint,
 amsmath,amssymb,
 aps,
 pra,
]{revtex4-2}
\usepackage[utf8]{inputenc}
\usepackage{graphicx}
\usepackage[colorlinks,linkcolor=blue,citecolor=blue]{hyperref}
\usepackage{breakurl}
\usepackage{array}
\usepackage{color,xcolor}
\usepackage{amsmath}
\usepackage{amsxtra}
\usepackage{amstext}
\usepackage{amssymb}
\usepackage{physics}
\usepackage{latexsym}
\usepackage{float}
\usepackage{xcolor}
\usepackage{booktabs}

\usepackage[english]{babel}
\usepackage[autostyle, english = american]{csquotes}
\MakeOuterQuote{"}

\newcommand{\gs} {3s^2 3p^5\ ^2P^{\circ}_{3/2}}
\newcommand{\gsp}{3s^2 3p^5\ ^2P^{\circ}_{1/2}}

\newcommand{\fived}{3s^2 3p^4({}^1D)5d\text{ }^2S_{1/2}}

\newcommand{\fourd}{3s^2 3p^4({}^3P)4d \text{ }^2D_{3/2}}
\newcommand{\fourdp}{3s^2 3p^4({}^3P)4d \text{ }^2D_{5/2}}

\newcommand{\threed}{3s^2 3p^4({}^1D)3d \text{ }^2P_{1/2}}
\newcommand{\threedp}{3s^2 3p^4({}^1D)3d \text{ }^2P_{3/2}}

\newcommand{\redmat}[3]{\left \langle #1 \middle|\middle| #2  \middle| \middle| #3 \right \rangle}

\graphicspath{./Figures/}
\begin{document} 

\title{Attosecond Transient Absorption Study of Coherent Hole Oscillation in Ar$^+$}

%\title{Coherent Hole Oscillation in Ar$^+$: Phases of Attosecond Transient Absorption Paths}

%\title{Attosecond Transient Absorption Study of Coherent Electron Dynamics in Ar$^+$: Analysis of Spin-orbit Wavepacket Phases}

%\title{Coherent Spin-Orbit Dynamics in Ar$^+$: Deviations from Perturbative Phase Relations in Attosecond Transient Absorption}

\author{Nisnat Chakraborty$^1$,
Miguel Alarc\'on$^1$, Michael P. McDonnell$^1$, Karl Hauser$^1$, Islam Shalaby$^1$, Dipayan Biswas$^1$, James K. Wood$^2$, Nikolay V. Golubev$^1$, Arvinder Sandhu$^{1,2,3}$}
\email{asandhu6@asu.edu}
\affiliation{$^1$Department of Physics, University of Arizona, Tucson, Arizona 85721, USA}
\affiliation{$^2$College of Optical Sciences, University of Arizona, Tucson, Arizona 85721, USA}
\affiliation{$^3$Department of Physics and CXFEL Laboratory, Arizona State University, Tempe, AZ, 85287, USA}
\date{\today}

\begin{abstract}
We report on the observation, characterization, and control of the electron dynamics in ionized argon atoms. We utilized an intense mid-infrared (MIR) pulse to create a coherent superposition of the spin-orbit split ground state of the ion. A weak extreme ultraviolet (XUV) pulse then probes the hole oscillation through time-resolved transient absorption spectroscopy. We investigated several 3\textit{p} to n\textit{d} transitions accessible with our XUV high harmonics which show a $23$~fs beat corresponding to the energy separation between the initially populated states. The experimental attosecond transient absorption signals for different pathways were simulated using detailed TDSE simulations and perturbative analytic calculations. The analysis of phase relations between the oscillations reveals important information about transition dipole moments in the system. In addition, we employed another strong MIR pulse to achieve transient control over the absorption by inducing Stark shifts of the states without affecting the electronic coherences. 
\end{abstract}

\maketitle

\section{Introduction}
Electronic motion in the valence orbitals of atomic and molecular systems occurs on timescales ranging from attoseconds to femtoseconds~\cite{Krausz2009}. Such ultrafast electron dynamics~\cite{Worner2017,Santra2011,Wirth2011} arise from the formation of coherent superpositions of electronic states. Understanding the origin and evolution of electronic coherences, as well as exploring ways to control them, can reveal fundamental properties of atomic and molecular systems and their response to external perturbations.

The tremendous developments in experimental laser science have made it possible to generate ultrashort light pulses capable of tracing the evolution of electronic coherences on their natural time scales. Table-top setups utilizing high-harmonic generation (HHG)~\cite{McPherson1987Studies,MFerray1988Multiple,Corkum1993Plasma} have become routinely available for the efficient production of sub-femtosecond pulses spanning the visible to soft X-ray spectral ranges~\cite{Calegari2016,Popmintchev2020}. Techniques such as attosecond streaking~\cite{Hentschel2001,Mairesse2005,Itatani2002}, reconstruction of attosecond beating by interference of two-photon transitions (RABBITT)~\cite{Paul2001,Mairesse2003}, high-harmonic spectroscopy~\cite{Rupenyan2013}, time-resolved photoelectron spectroscopy, attosecond transient absorption spectroscopy (ATAS) have been used to measure the electron dynamics in a large variety of atomic and molecular systems\cite{Shalaby2022,YanezPagans2025,Timmers2014,Biswas2024} as well as  in the condensed matter~\cite{Lucchini2016,Gannan2025,deRoulet2024}.

%Although isolated atoms are among the simplest and most extensively studied systems, they also provide a convenient platform for exploring subtle aspects of the electron motion and for developing reliable methods to control the electronic coherences. Among other advantages, atomic systems are known to maintain coherence for a long time due to a limited number of internal degrees of freedom which otherwise can lead to a fast decoherence, as it is typically the case for molecules~\cite{Scheidegger2022}. Atoms are also easy to isolate from the environment and minimize their interactions with each other, for example in case of noble gases, which is crucial for studying the electron motion and its experimental observations. Furthermore, the electronic structure of atoms can be manipulated via optical and magnetic fields in a wide range of different regimes from coherent population transfer and strong field Rabi oscillations control to transient control of energy levels via Stark and Zeeman effects in both ultrafast and long-term manner. All together these unique properties of atomic systems made isolated atoms indispensable in a variety of cutting-edge technologies such as quantum computing~\cite{YagoMalo2024} and communications~\cite{Saha2025}, precise atomic clocks~\cite{Oskay2006}, and tests of fundamental aspects of physics~\cite{Panda2024}.

Here we use a few-cycle mid-infrared (MIR) laser pulse to drive the strong field ionization (SFI) in a gas of argon atoms, creating a coherent superposition of the lowest spin-orbit-coupled electronic states of the ion. The resulting electronic coherence is probed with an extreme-ultraviolet (XUV) attosecond pulse train. We measure the ATAS signal in the energy range of 21--28~eV which corresponds to the 17th, 19th, and 21st harmonics of the MIR laser. We trace the oscillations, intensities, and phases of nine distinct absorption lines and highlight the non-intuitive relationships between different ATAS signals.

The ATAS approach used in our experiment has been previously employed to study valence-shell wave packet dynamics in krypton~\cite{Goulielmakis2010} and  xenon~\cite{Kobayashi2018}. Its extension to molecular systems has the potential for capturing charge migration on attosecond timescales~\cite{Matselyukh2022, Golubev2021}. Yet, a detailed analysis of the relative phases and intensities of various absorption lines in the ATAS spectra has not been reported so far. Here, we  establish the correspondence between the measured spectra and the properties of the coherent superposition of electronic states, as well as values of the transition dipole moments connecting initial and final states.

To verify our interpretations, we perform a series of fully quantum mechanical simulations of the ATAS signal solving the time-dependent Schr\"odinger equation (TDSE) numerically with explicit consideration of the interactions between the system and the applied external fields. We compare the results of our simulations with the spectra obtained within the perturbation theory framework~\cite{Santra2011}. While numerical integration of the TDSE is more versatile, enabling the calculation of the absorption spectra in the presence of the control fields or when the pump and the probe overlap, we still find approximate analytical perturbative expressions valuable. Their analysis helps unravel the essence of the physical mechanisms that govern the dynamics of the observation. For instance, using such expressions enables us to present simple mathematical expressions to explain the phase relations between different oscillations. 
%
%We demonstrate that although the straightforward propagation of the wavefunction achieved by numerical solution of the TDSE is more universal since it permits simulations of the absorption spectra in the presence of control fields or when pump and probe lasers overlap, the usage of less accurate but analytic perturbative approach helps to unravel the essence of physical mechanisms underlying the time-dependent absorption spectroscopy. We present simple mathematical expressions explaining the reasons for the observed phase offsets in oscillations of the transient absorption lines.
%
\begin{center}
\begin{figure*}[t]
\includegraphics[width=\textwidth]{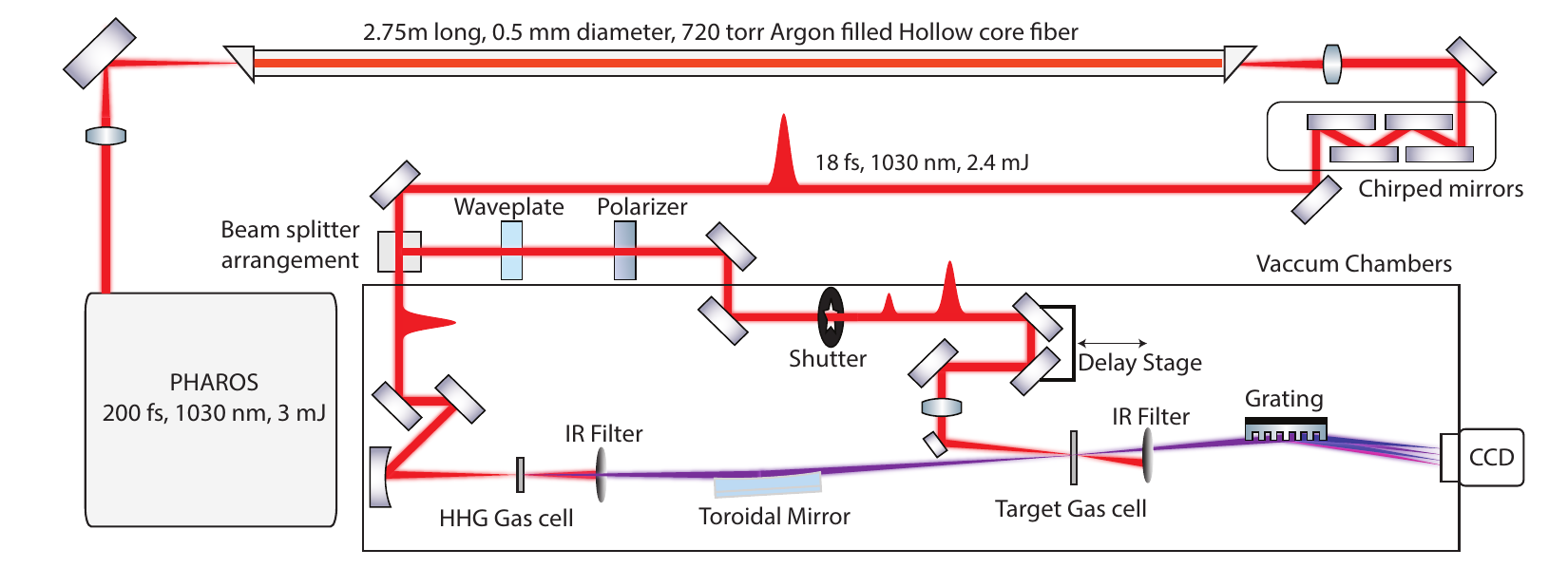}
\caption{Schematic of the transient absorption spectroscopy setup. A commercial Yb-based laser, coupled with a 2.75~m long hollow-core fiber and a chirped mirror compressor, produces 2.4~mJ, 18~fs pulses centered at 1030~nm. These MIR pulses are subsequently split into three arms---pump (MIR), probe (XUV), and control (MIR)---with a special beam splitter arrangement, using a 60:40 beam splitter and a pair of holey mirrors. All three beams are ultimately recombined at the target gas cell, where the transmitted XUV pulse is directed to a spectrometer for detection.}
\label{fig:Setup}
\end{figure*}
\end{center}

Beyond observation and characterization of the ATAS signals, we investigate the active laser control of the electronic processes by applying a delayed strong MIR field. We demonstrate the possibility to induce the transient Stark shifts on the absorption lines without affecting the coherent superposition of the ionic states. This control capability lays the groundwork for quantum non-demolition control techniques enabling manipulations of quantum states without destroying their coherence which is critical for quantum error correction and sensing.

The paper is organized as follows. In Sec.~\ref{sec:exp}, we present the detailed description of our experimental setup. Section~\ref{sec:theory} contains the description of the computational and theoretical schemes. We discuss the observed coherent electron dynamics in Sec.~\ref{sec:results_two_pulse} and present our scheme for transient control of the system in Sec.~\ref{sec:results_control}. We summarize our results and conclude the paper in Sec.~\ref{sec:concl}.

\section{Experiment}
\label{sec:exp}
Our experimental system, depicted in Fig.~\ref{fig:Setup}, utilizes a commercially available ytterbium based laser system (Pharos, LightConversion USA) which produces 3~mJ, 200~fs pulses at a central wavelength of 1030~nm. The pulses are focused into a 2.75~m long hollow-core fiber filled with argon which is pressurized to 720~Torr, producing a MIR supercontinuum in the spectral range of about 950~nm to 1100~nm. The broad MIR output is temporally compressed by four chirped mirrors, generating 18~fs pulses with a pulse energy of 2.4~mJ~\cite{Shalaby2025}.

The MIR pulse is subsequently divided into three parts. The first 60$\%$ is directed into a gas cell backed by 50~torr xenon, where it drives the HHG. The HHG process is phase-matched to achieve efficient generation of XUV in the energy range of 21--28~eV, mainly through the 17th, 19th, and 21st harmonics of the MIR beam of wavelength 1030~nm. This spectral window gives access to the $3d$, $4d$, and $5d$ spin-orbit split states of atomic argon ion, as we discuss in details in Sec.~\ref{sec:theory} of this paper. The XUV pulse train co-propagates with the residual MIR field. The MIR radiation is subsequently removed by a 200~nm thick aluminum filter and the filtered XUV beam is focused to a spot size of 70~$\mu$m using a toroidal mirror with a 1~m focal length. At the focus XUV interacts with an effusive argon gas jet at a backing pressure of 10~Torr. Post interaction, the XUV beam proceeds to a spectrometer consisting of a curved variable line space diffraction grating, which spectrally resolves and images the XUV transmission from the focal region onto an X-ray charge-coupled device (CCD) detector.

Out of the remaining 40$\%$ of the MIR beam, 37$\%$ is employed as a non-collinear pump pulse to ionize the argon atoms by strong-field (tunnel) ionization. The resulting ions Ar$^{+}$ and any residual neutrals are probed with the attosecond XUV pulse train. Any scattering of the non-collinear MIR beam along the XUV propagation direction is eliminated by a second aluminum filter. Temporal delay between the MIR pump and XUV probe arm is introduced using a precision delay stage that controls the optical path length of the MIR beam. This delay is systematically scanned to capture the time-resolved evolution of the coherent electron dynamics. At each delay position, the transmitted XUV spectrum is recorded using the CCD. The delay-dependent absorption features are obtained by averaging over 120 alternate probe-only (XUV) and pump–probe (XUV + MIR) spectra, with each frame captured at an exposure time of 0.5~seconds.

The transient absorption signal is quantified by the change in optical density, $\Delta$OD($\tau$), defined as
\begin{equation}
\label{eq:OD}
    \Delta \text{OD}(\tau) = -\log\left( \frac{I_{\text{XUV+MIR}}(\tau)}{I_{\text{XUV}}} \right),
\end{equation}
where $I_{\text{XUV+MIR}}(\tau)$  and  $I_{\text{XUV}}$  denote the transmitted XUV intensities through the gas target with and without the MIR pump, respectively, as a function of the pump-probe delay $\tau$.

Lastly, the remaining 3$\%$ of the MIR pulse is reserved as a control field, introduced at a fixed delay of 750~fs with respect to the pump pulse. This control pulse is tuned to be strong enough to induce the Stark shift of energy levels in the system yet being too weak to cause any noticeable population transfer or other strong-field effects on the initially created electronic coherences.

\begin{table}[t]
\caption{Dipole-allowed electronic transitions in Ar$^+$ in the energy range from 21 to 28~eV. The energies and electronic configurations are taken from the NIST Atomic Spectra Database~\cite{NIST_ASD_2024}. We assign a symbolic notation $\nu_i$ to each of the transitions in order to simplify the corresponding discussion in the main text.}
\begin{ruledtabular}
\begin{tabular}{cccc}Energy (eV) & Initial State & Final State & Symbol \\[5mm]
\midrule
 26.8900 & $\gs$   & $\fived$  & $\nu_1$\\[3mm]
 26.7125 & $\gsp$  & $\fived$  & $\nu_2$\\[3mm]
 23.8933 & $\gs$   & $\fourd$  & $\nu_3$\\[3mm]
 23.7158 & $\gsp$  & $\fourd$  & $\nu_4$\\[3mm]
 23.8740 & $\gs$   & $\fourdp$ & $\nu_5$\\[3mm]
 21.6750 & $\gs$   & $\threed$ & $\nu_6$\\[3mm]
 21.4975 & $\gsp$  & $\threed$ & $\nu_7$\\[3mm]
 21.6241 & $\gs$   & $\threedp$& $\nu_8$\\[3mm]
 21.4466 & $\gsp$  & $\threedp$& $\nu_9$\\[3mm]
\end{tabular}
\end{ruledtabular}
\label{tab:lines}
\end{table}

\section{Theory}
\label{sec:theory}

\subsection{Description of the system}
The ground electronic configuration of Ar$^+$ is $3s^2 3p^5$. According to the NIST Atomic Spectra Database~\cite{NIST_ASD_2024}, the ground electronic state experiences a considerable spin-orbit splitting leading to the creation of four degenerate electronic states of $^2P^{\circ}_{3/2}$ character and two degenerate states of $^2P^{\circ}_{1/2}$ character separated by 0.177~eV from each other. The next ionic state of Ar$^+$ has $3s 3p^6\ (^2S_{1/2})$ electronic configuration and is almost 13.5~eV above the ground state. This large energy separations implies that following SFI by the MIR, the resulting ionic wave packet will predominantly be composed of the six lower states of the cation. We assume an equal linear superposition of those six initial states as there no reason to favor any of the magnetic quantum numbers due to the symmetry of the experiment. 
%We leave a more detailed simulations of the interaction of the neutral ion with an ionizing laser pulse for a future study capable of better describing the ionization process.

\begin{figure}[tp]
\includegraphics[width=\columnwidth]{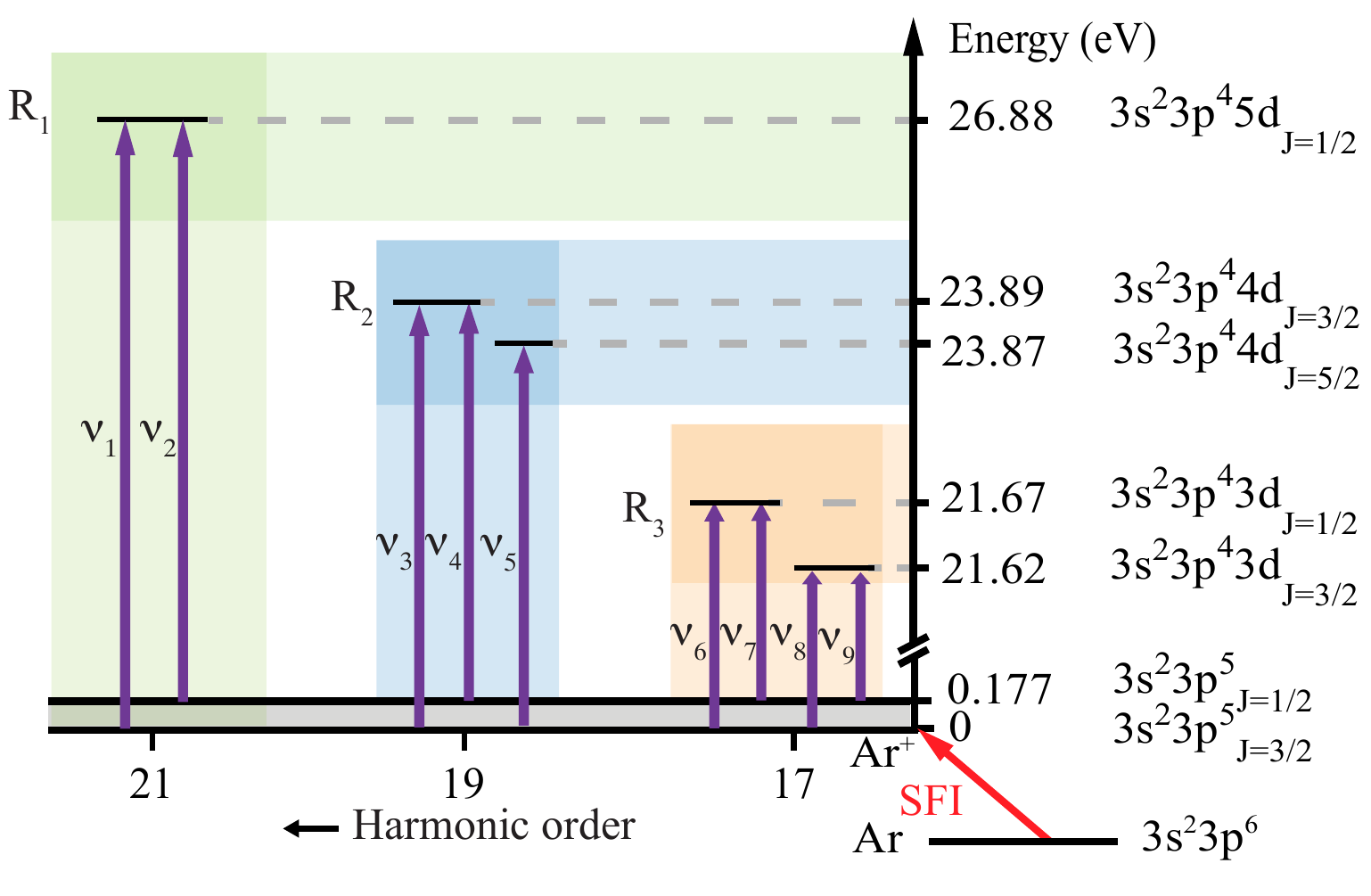}
\caption{Schematic representation of the experiment for studying the electronic dynamics in Ar$^+$. An intense mid-infrared (MIR) laser pulse creates electronic coherences in the spin-orbit split $3s^{2}3p^{5}_{J=3/2}$ and $3s^{2}3p^{5}_{J=1/2}$ electronic states of the cation. A weak extreme ultraviolet (XUV) laser pulse is used to probe the dynamics in the system via transmission of the 21$^{\text{st}}$ (region R$_1$, highlighted in green), 19$^{\text{th}}$ (region R$_2$, highlighted in blue), and 17$^{\text{th}}$ (region R$_3$, highlighted in orange) harmonics. %A moderately intense MIR control pulse is used to achieve a transient control over the electron dynamics in the system without destroying the created electronic coherences.
}
\label{fig:levels}
\end{figure}

In the energy window between 21 and 28~eV covered by the XUV probe pulse, the initially populated electronic states can be coupled to a large number of final electronic states. For simplicity we focus on the dipole-allowed transitions leading to a total of nine distinct absorption lines, which are listed in Tab.~\ref{tab:lines} and schematically shown in Fig.~\ref{fig:levels}. The electronic configurations of the corresponding states were obtained from the NIST database and used to compute the transition dipole moments as detailed in the Appendix~\ref{appdx:DM}.

Since the reference for quantifying the  optical density change in the experiment is the transmission of XUV pulse only through neutral sample (denominator of Eq.~(\ref{eq:OD})), we also considered several absorption lines of the neutral system. Specifically, we included $3s 3p^6 4p (^1P)$, $3s 3p^6 5s (^1S)$ and $3s 3p^6 3d (^1D)$ autoionizing electronic states of the neutral atom to our model. While these states have no impact on the electron dynamics in the ionic system, the neutral transitions can affect the absorption spectrum and  get mixed in with the ionic absorption signals, especially in the MIR control regime as we discuss in the Sec.~\ref{sec:results_control}.

\subsection{Attosecond transient absorption spectroscopy}
\label{sec:ATAS_theory}
Application of an external electric field to a quantum system can cause population transfer between energy levels of that system. The probability of this process is maximized when the  applied photon energy is resonant with the energy gap between the involved states. By measuring the transmission of the light pulses through a sample, one can observe a depletion in the field intensity at frequencies corresponding to specific resonances. The intensities of the observed absorption lines are proportional to the square of the electric dipole moments connecting the initial and final states.

If the system is in a quantum superposition when interacting with a short laser pulse, due to the interferences between population transfer pathways, intensity of the absorption lines oscillates as a function of the pulse arrival time. A necessary condition for oscillations to exist, beyond the presence of the corresponding electronic coherences, is the coupling between the states that form the initial wave packet and the final states. The frequency of these oscillations is determined by the energy difference between the states of the initial wave packet. Thus, if one observes an absorption line oscillating in time, the frequency spectrum provides important information about the pairs of states that couple to the probed final state. The derivations and the detailed analysis of this time-dependent version of absorption spectroscopy, also known as ATAS in the literature~\cite{Santra2011,Golubev2021}, have attracted considerable interest in recent years. 

For a weak enough probe pulse, the ATAS cross-section can be obtained analytically employing first-order perturbation theory directly to the time-dependent initial state of the system (see, e.g., Refs.~\cite{Santra2011,Golubev2021}). Assuming the probe field has a Gaussian shape and neglecting certain terms (see Appendix~\ref{appdx:gauss} for the detailed derivation), the absorption cross-section can be expressed as
\begin{widetext}
\begin{equation}
\label{eq:TA}
    \sigma(\omega,\tau) = \frac{4\pi \omega}{c} \text{Im} 
        \sum_{k,j} c_k^* c_j e^{i\tau(\epsilon_j-\epsilon_k)}
        e^{-\frac{\gamma^2}{4}(\epsilon_k-\epsilon_j)^2} e^{-\frac{\gamma^2}{2}(\omega-\omega_0)(\epsilon_j-\epsilon_k)}
        \sum_f \mu_{kf} \mu_{fj}
    \left[
         \frac{1}{\tilde{\epsilon}_f  -\epsilon_k-\omega}
        +\frac{1}{\tilde{\epsilon}_f^*-\epsilon_j+\omega}
    \right],
\end{equation}
\end{widetext}
where $\omega$ and $\tau$ denote the spectral frequency and time delay, respectively, at which the absorption is evaluated, $c$ is the speed of light in vacuum, $\omega_0$ is the photon energy of the applied field and $\gamma = \sqrt{2\ln2}/\text{FWHM}$ is the parameter related to the full width at half maximum (FWHM) of the Gaussian pulse. Sub-indices $k$ and $j$ run over the states forming the initial wave packet with, in general complex, coefficients $c_k$ and $c_j$ and energies $\epsilon_k$ and $\epsilon_j$, respectively. Index $f$ iterates over final states with complex energies $\tilde{\epsilon}_f=\epsilon_f-i(\Gamma_f/2)$, where $\Gamma_f$ is the parameter responsible for the finite lifetime of the state, which are coupled to the initial ones via dipole matrix elements $\mu_{kf}$ and $\mu_{fj}$. 

%Move this one somewhere in the appendix
%The obtained expression is different from a comparable one obtained assuming the probe pulse is infinitely short in time by the presence of $e^{-\frac{\gamma^2}{4}(\epsilon_k-\epsilon_j)^2} e^{-\frac{\gamma^2}{2}(\omega-\omega_0)(\epsilon_j-\epsilon_k)}$ factors accounting for the final bandwidth of the utilized pulse.

According to the Eq.~(\ref{eq:TA}), the absorption lines will show up at the vicinities of photon energies matching the energy difference between the initial and final states (terms in square brackets). The energy profiles of these lines are determined by the product of a Lorentzian, due to the finite life time of the state, with a Gaussian, due to the finite duration of the applied field. Observing a line in the ATAS spectrum depends on multiple factors. Namely, on the populations of the initial states entering Eq.~(\ref{eq:TA}) via $c_k$ and $c_j$, on the corresponding dipole moments $\mu_{kf}$ and $\mu_{fj}$ connecting the initial and final states with each other, and also on the photon energy $\omega_0$ and width $\gamma$ of the applied electric field. The oscillating behavior of the lines is produced by the $e^{i\tau(\epsilon_j-\epsilon_k)}$ terms which are responsible for the electronic coherences created in the system. The phases of these oscillations depend on how the initial wave packet is created. Note, however, that these phases are ``global'' which means that they will be identical across all the lines in the spectra connecting to the same initial states.

While the above description of the time-dependent transient absorption spectroscopy is straightforward, the ATAS spectra measured in a realistic system could be quite complex due to the interplay of multiple effects contributing to Eq.~(\ref{eq:TA}). In Sec.~\ref{sec:results_two_pulse}, we present various limiting cases arising from Eq.~(\ref{eq:TA}) and use them to interpret the results of our experiments by disentangling various effects in the absorption spectra of the argon cation. To simplify the discussion even further, we assume that the probe field has a uniform energy profile thus setting $\gamma=0$. Furthermore, we introduce a global phase $\varphi$, arising as a result of the SFI, such that $c_k^* c_j = |c_k| |c_j| e^{i \varphi}$, which we will adjust to shift the overall signal to better match the experimentally measured spectra.

\subsection{Numerical simulations}
While the analytic formula for the absorption cross-section in Eq.~(\ref{eq:TA}) gives access to a simple and intuitive interpretation of the physics underlying the ATAS measurements, it is limited to analyzing transitions in the ``field-free'' regime only, when the system does not experience action of other electric fields except the probe. Furthermore, the linear-response perturbation theory used in the derivation of Eq.~(\ref{eq:TA}) cannot account for interferences of various pathways potentially involved in the population transfer in systems with rather dense electronic structure such as Ar$^+$.

To circumvent the aforementioned issues and go beyond the limitations imposed by the perturbative treatment, we simulate the electron dynamics and the action of the control and the probe fields on the system by numerically solving the TDSE using the split operator method~\cite{Plunkett2022,Tarana2012}. The Hilbert space in which the TDSE is propagated is composed from all the spin-orbit coupled states of Ar$^+$ present in the NIST database~\cite{NIST_ASD_2024} (3238 states in total). Assuming the validity of the dipole approximation, we constructed the total Hamiltonian of the system as
\begin{equation}
\label{eq:ham}
    \hat{H}(t,\tau) = \hat{H}_0 - \left[ \vec{V}_{\text{MIR}}(t-t_{\text{MIR}}) + \vec{V}_{\text{XUV}}(t-\tau) \right] \cdot \hat{\vec{D}},
\end{equation}
where the field-free Hamiltonian $\hat{H}_0$ is diagonal in this representation and is directly composed from the energies taken from the NIST, while the non-diagonal dipole operator $\hat{\vec{D}}$ is computed based on the electronic configurations of the corresponding states imported from NIST using the prescription detailed in the Appendix~\ref{appdx:DM}. The control $\vec{V}_{\text{MIR}}(t)$ and probe $\vec{V}_{\text{XUV}}(t)$ fields are assumed to have Gaussian envelopes
\begin{equation}
    \vec{V}(t) = \vec{\epsilon} \sqrt{I} e^{-( \sqrt{2\ln 2} t/\text{FWMH} )^2 } \cos(\omega t),
\end{equation}
where the unit vector $\hat{\epsilon}$ denotes the polarization vector of the field, $I$ is the field intensity, FWHM parameter controls the duration of the field, and $\omega$ is the photon energy. The corresponding parameters of MIR and XUV fields used in our simulations are given in Tab.~\ref{tab:field_params}. Note that the XUV field is, in principle, composed from the three harmonics, each of which is described by a separate Gaussian centered at $\tau$ time delay. The arrival time $t_{\text{MIR}}$ of the control field is kept fixed at 750~fs, as specified by the design of the experimental setup. We assumed all the fields are linearly polarized in the $z$-direction, being aligned with the ionizing pulse, and that they have zero carrier-envelope phase (CEP).

\begin{table}[t]
\caption{Parameters of MIR control and XUV probe fields used in our numerical simulations}
\begin{ruledtabular}
\begin{tabular}{cccc}
 Field & $I$ (TW/cm$^2$) &  FWHM (fs)  & $\omega$ (eV) \\
\midrule
 MIR & 20 & 18 & 1.2526 \\
 XUV, 17$^{\text{th}}$ & 5$\times 10^{-4}$& 3 & 21.2947\\
 XUV, 19$^{\text{th}}$ & 5$\times 10^{-4}$& 3 & 23.8 \\
 XUV, 21$^{\text{st}}$ & 5$\times 10^{-4}$& 3 & 26.3052 \\
\end{tabular}
\end{ruledtabular}
\label{tab:field_params}
\end{table}

\begin{figure*}[t]
\includegraphics[width=\linewidth]{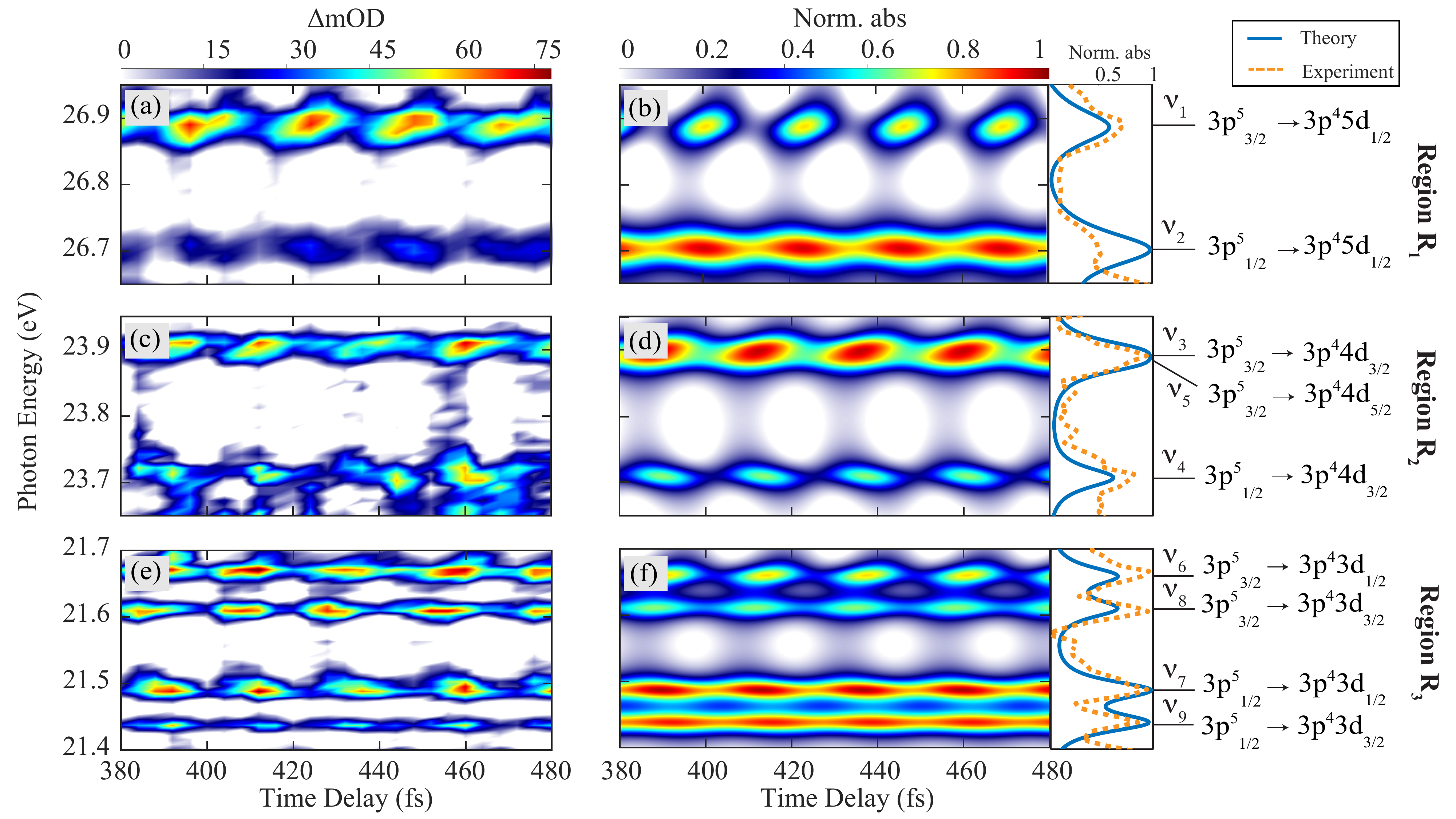}
\caption{Experimentally measured (left column) and theoretically simulated (right column) time-resolved transient absorption spectrograms of argon cation created via strong field ionization process. Horizontal axes represent the time delay between the MIR pump and XUV probe pulses. Vertical axes correspond to the measured photon energy of the XUV laser. The color contours depict the observed optical density which varies as a function of the time delay and the photon energy. Panels (a) and (b) depict the energy region $R_1$ (see Fig.~\ref{fig:levels}), which contains the transitions from the spin-orbit split $3s^{2}3p^{5}_{J=3/2}$ and $3s^{2}3p^{5}_{J=1/2}$ initial states to the final $3s^{2}3p^{4}5d_{J=1/2}$ electronically excited state of the cation. Similarly, panels (c) and (d) correspond to the region $R_2$, showcasing transitions to final $3s^{2}3p^{4}4d_{J=3/2}$ and $3s^{2}3p^{4}4d_{J=5/2}$ states. Finally, panels (e) and (f) illustrate region $R_3$, capturing the dynamics associated with transitions to $3s^{2}3p^{4}3d_{J=1/2}$ and $3s^{2}3p^{4}3d_{J=3/2}$ final states.}
\label{fig:full_spec}
\end{figure*}

We calculate the change in the optical density following the procedure described in Ref.~\cite{Wu2016}. Accordingly, we use the following expression for the absorption cross-section
\begin{equation}
    \sigma(\omega,\tau) = \frac{4 \pi \omega}{c} \text{Im}
        \left(
            \frac{\tilde{d}(\omega,\tau)}
            {\tilde{V}_{\text{MIR}}(\omega) + \tilde{V}_{\text{XUV}}(\omega)}
        \right),
\end{equation}
where $c$ is the speed of light in vacuum, $\tilde{V}_{\text{MIR}}(\omega)$ and $\tilde{V}_{\text{XUV}}(\omega)$ denote the Fourier transforms of MIR and XUV fields, respectively, and $\tilde{d}(\omega,\tau)$ is the Fourier transform of the time-dependent dipole moment of the system:
\begin{equation}
    d(t,\tau) = \bra{\psi(t,\tau)} \hat{D}_z \ket{\psi(t,\tau)}.
\end{equation}
The evolution of the electronic wavefunction $\ket{\psi(t,\tau)}$ in time is computed under the action of the Hamiltonian in Eq.~(\ref{eq:ham}). Depending on the arrival time $\tau$ of the probe field, the evolution of the system will, in general, change, causing a modification in the corresponding dipole moment. We, therefore, perform a scan of the dynamics in the system by varying the pump-probe delay $\tau$ to mimic the conditions of the experiment. The time grid used to propagate the wave function spans from $0$ to $1000$ fs using a time step dependent on the XUV frequency close to $0.01$ fs. To numerically converge the Fourier transform of the computed dipole moment $d(t,\tau)$, we apply the exponential decay filter with a characteristic time scale of 40~fs, thus mimicking the finite lifetime of the electronic states reached by the probe pulse.

To ensure the validity of our calculations and to speed up the corresponding data analysis, we conducted a series of simulations varying the number of electronic states present in our Hamiltonian as well as the composition of the probe field. We used the complete system with all the electronic states and all the three XUV harmonics as a reference and compared it with the results obtained using reduced system where the Hilbert space was limited to the ionic ground states, the states approximately in resonance with the harmonics, and those within bandwidth of the MIR, switching on and off the presence of the corresponding harmonics. We, therefore, explored the importance of high-lying and neighboring electronic states in the  system, as well as possible interference between various pathways initiated by the control and probe fields. Since our reduced model produced the absorption cross-section visually indistinguishable from the one obtained with the full system while demonstrating considerable computational advantages, all the results shown later in this work come from that reduced effective model.

\section{Results and Discussion}
Below we present the measured ATAS signals and discuss various aspects of the recorded spectra. To better analyze and understand the physics that leads to each one of these observed features we will split the data in two sets. First, in Sec.~\ref{sec:results_two_pulse} we will discuss the data obtained using the ionizing and the probe pulses only, thus studying the electronic coherences and the connections between the initial and final states in the system. Second, in Sec.~\ref{sec:results_control} we explore the effects of a third strong MIR control pulse on the system and investigate the changes in the measured spectra during and after the action of the applied field.

\subsection{Two pulse experiment}
\label{sec:results_two_pulse}
Our experimental ATAS signal is shown in the left column of Fig.~\ref{fig:full_spec}. To facilitate the analysis, we have split the spectrogram in three regions (see Fig.~\ref{fig:levels}) and used the symbols from Tab.~\ref{tab:lines} to refer to each one of the absorption lines. Region $R_1$, accessed by the 21st harmonic of our XUV pulse, covers energy range between 26.50 and 26.98~eV where two lines corresponding to transitions $\nu_1$ and $\nu_2$ are present. Similarly, region $R_2$, between $23.504$ and $23.98$~eV covered by the 19th harmonic, includes transitions $\nu_3$, $\nu_4$ and $\nu_5$; and region $R_3$, between $21.28$ to $21.76$ eV covered by the 17th harmonic, contains the remaining $\nu_6$, $\nu_7$, $\nu_8$ and $\nu_9$ lines. We clearly observe all nine transitions listed in Tab.~\ref{tab:lines}. The assignments of the absorption lines along with the results of our theoretical simulations are shown in the right column of Fig.~\ref{fig:full_spec}.

The first thing to notice in the spectra depicted in Fig.~\ref{fig:full_spec} is that all the observed absorption lines oscillate with a period of 23~fs which corresponds to the 0.177~eV spin-orbit energy splitting of the ionic ground state of argon, which is expected based on the results of a similar experiment~\cite{Goulielmakis2010} performed in krypton ions. Nonetheless, time and energy resolution in our measurements permits the analysis of phases of the observed oscillations.

Let us start by considering the region $R_1$ in the spectra where only a single final state is present in the system (see Fig.~\ref{fig:levels}). Accordingly, two initial states coupled to a single final state show up in the spectrogram in the form of two oscillating lines $\nu_1$ and $\nu_2$. Indeed, the measured absorption signal, shown in panel (a) of Fig.~\ref{fig:full_spec}, demonstrates this expected behavior. Interestingly, while both absorption lines are observed to be oscillating in phase, the changes of their individual phases as functions of photon energy are found to behave in the opposite way with respect to each other. As one can see, the slope of oscillatory features in the spectrogram is positive for the upper line and negative for the lower line.

To explain this feature of the time-dependent absorption spectra, we utilize a general perturbative formalism presented in Sec.~\ref{sec:ATAS_theory}. Assuming that the system is composed of two initial and one final state, rearranging terms in Eq.~(\ref{eq:TA}), and using the notation $\epsilon_{12}=\epsilon_1 - \epsilon_2$, the oscillatory part of the cross-section can be written as
\begin{widetext}
\begin{equation}
\label{eq:twolines}
    \sigma_{\text{osc}}(\omega,\tau) = \frac{4\pi \omega}{c} |c_1| |c_2| \mu_{1f} \mu_{f2}
    \left\{
        L_{f1}(\omega) \sin\left[\epsilon_{12}\tau+\phi_{f1}(\omega)\right]
       -L_{f2}(\omega) \sin\left[\epsilon_{12}\tau-\phi_{f2}(\omega)\right]
    \right\},
\end{equation}
\end{widetext}
where we introduced the frequency dependent amplitude
\begin{equation}
    L_{fk}(\omega) = \sqrt{\frac{1}{(\varepsilon_f-\epsilon_k-\omega)^2+\Gamma_f^2/4}},
\end{equation}
and phase
\begin{equation}
     \phi_{fk}(\omega)=\arctan \frac{\Gamma_f/2}{\varepsilon_f-\epsilon_k-\omega},
\end{equation}
which determine the energy profiles and oscillatory behavior, respectively, of the absorption lines. Note that the prefactor in front of the curly brackets ``$\{...\}$'' in Eq.~(\ref{eq:twolines}) is common for both transitions and do not contribute to potential phase offset between the lines. The expression inside the curly brackets shows that if the initial states have a larger splitting $\epsilon_{12}$, and hence a larger energy separation of two lines in the spectra, than the width $\Gamma_f$ of $L_{fj}$ profiles, then the two oscillating lines are well separated and their phases do not interfere with each other. Measuring the spectra at the peaks of $L_{f1}$ and $L_{f2}$, where the corresponding phase factors $\phi_{f1}$ and $\phi_{f2}$ are both $\pi/2$, transforms both sine terms to identical cosines: $\sin(\epsilon_{12}\tau + \pi/2)=\cos(\epsilon_{12}\tau)$ and $-\sin(\epsilon_{12}\tau - \pi/2)=\cos(\epsilon_{12}\tau)$, oscillating with the same frequency $\epsilon_{12}$ and having identical phases. At the same time, the behavior of the absorption lines away from the resonances, while still oscillating with identical frequency, differ from each other due to the opposite signs in front of the corresponding sine terms. Accordingly, the opposite variations of the phases of these two lines along their energy profiles explain the different slope of the corresponding oscillatory features. The calculated absorption cross section, depicted in panel (b) of Fig.~\ref{fig:full_spec}, demonstrates an excellent agreement with the measured spectrograms highlighting the validity of our interpretations. 

We note the possibility for the existence of a more complicated scenario when the two absorption lines have considerable overlap between each other due to a small separation between the initial states. In this case, although still having the same frequency, the phases at either of the absorption peaks will interfere being affected by the tails of the neighboring absorption lines. While this exact mechanism has not been observed in our study, a similar effect of overlapping absorption lines, which arise due to the close proximity of final states, plays an essential role in the absorption spectra as we explain next.

One can expect that the region $R_2$, shown in panels (c) and (d) of Fig.~\ref{fig:full_spec}, should, in principle, contain more sophisticated features than region $R_1$. This is due to the presence of an additional absorption line $\nu_5$ which is coupled to one of the initial states of the system (see Fig.~\ref{fig:levels}). At the first glance, however, this line should not contribute much to the absorption spectra. Indeed, in the case when only one of the initial states is coupled to a final state, so that the other transition dipole moment is zero, the oscillating part of the absorption cross-section, Eq.~(\ref{eq:twolines}), should vanish completely. Yet, due to the overlap of $\nu_3$ and $\nu_5$ transitions, the actual energy profile and the phase behavior of the upper absorption line in the region $R_2$ is more intricate than the one observed in the region $R_1$. However, this effect is minor and can barely be observed in the experimental data. Accordingly, the overall structure of the $R_2$ region is nearly indistinguishable from that of $R_1$.

\begin{figure}[t]
\includegraphics[width=\linewidth]{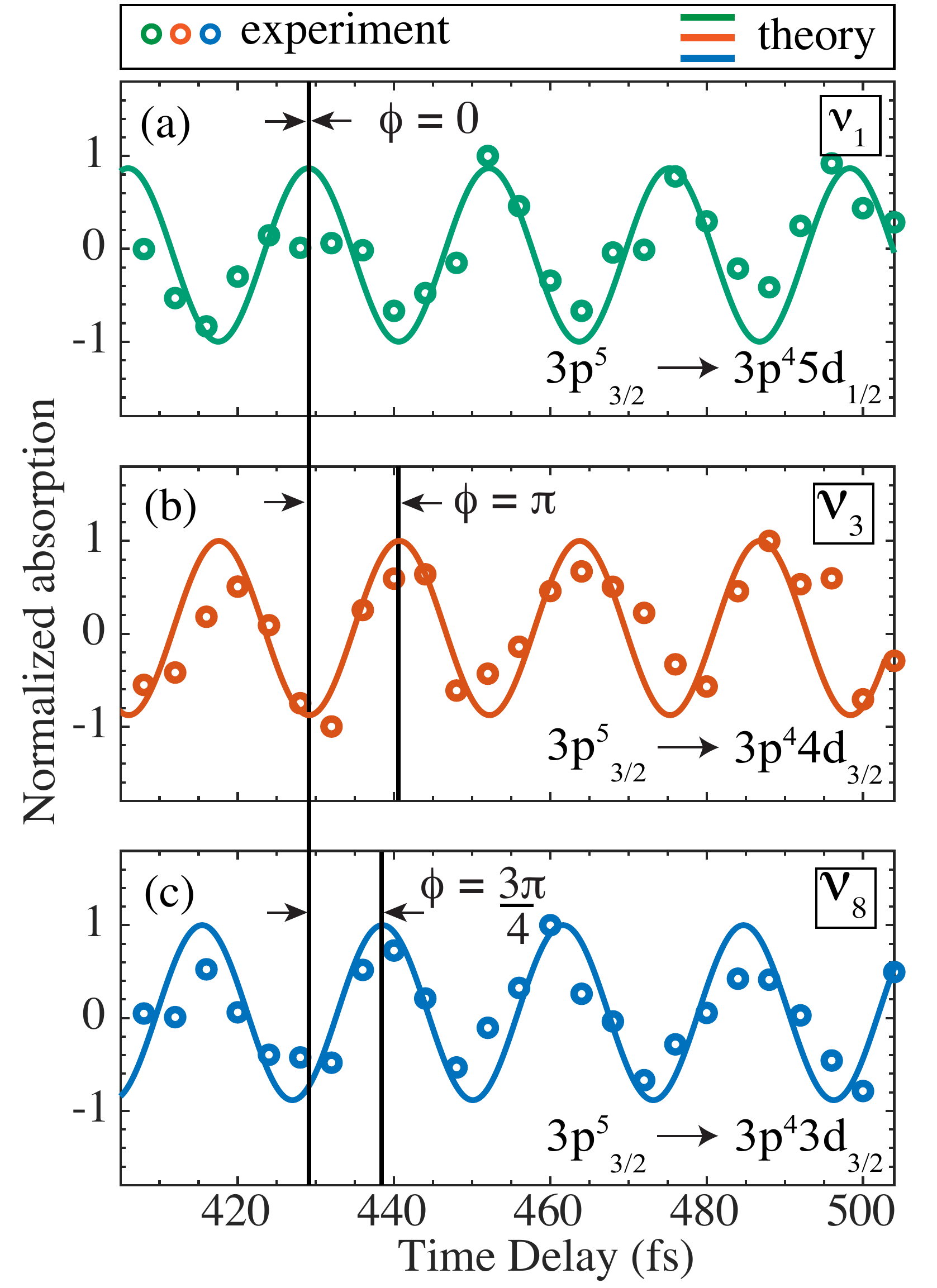}
\caption{Lineouts from Fig.~\ref{fig:full_spec} at energies corresponding to the transitions from $3s^2 3p^5_{J=3/2}$ state to $3s^2 3p^4\, \text{n}d$ manifold, showing experimental data (dots) and theoretical simulations (solid lines). Phase flip between the $\nu_1$ and $\nu_3$ lines is attributed to difference in the sign of the transition dipole moments. The $3\pi/4$ phase offset of the $\nu_8$ line stems from the overlap of absorption lines in the vicinity of $\nu_8$ transition. To make the phase extraction clearer, the DC component was removed from each of the lineouts and then normalized to a unit amplitude range.}
\label{fig:lineout1}
\end{figure}

Nonetheless, a major distinction between the absorption lines present in the regions $R_1$ and $R_2$ is a noticeable difference in the phases of the oscillations in two regions. Panels (a) and (b) of Fig.~\ref{fig:lineout1} depict the normalized lineouts of one absorption line from each region, i.e. $\nu_1$ and $\nu_3$ from $R_1$ and $R_2$ respectively, extracted at the corresponding resonance frequencies. As one can see, these two absorption lines oscillate $\pi$ radians out of phase with respect to each other. To understand the nature of this effect, let us consider the absorption cross-section for the case where the two initial states are coupled to two different final states:
\begin{widetext}
\begin{equation}
\label{eq:fourlines}
\begin{aligned}
    \sigma_{\text{osc}} (\omega,\tau) = \frac{4\pi \omega}{c} |c_1| |c_2|
    \Bigl(
        & \mu_{1f_{1}} \mu_{f_{1}2} 
        \left\{
            L_{f_11}(\omega) \sin[\epsilon_{12}\tau+\phi_{f_11}(\omega)] 
           -L_{f_12}(\omega) \sin[\epsilon_{12}\tau-\phi_{f_12}(\omega)]
        \right\} \\
    +  & \mu_{1f_{2}} \mu_{f_{2}2}
        \left\{
            L_{f_21}(\omega) \sin[\epsilon_{12}\tau+\phi_{f_21}(\omega)]
           -L_{f_22}(\omega) \sin[\epsilon_{12}\tau-\phi_{f_22}(\omega)]
        \right\}
    \Bigl).
\end{aligned}
\end{equation}
\end{widetext}

The amplitudes of the peaks $L_{f_{1}1}$/$L_{f_{1}2}$, and $L_{f_{2}1}$/$L_{f_{2}2}$ are controlled by the products of the dipole matrix elements $\mu_{1f_{1}} \mu_{f_{1}2}$ and $\mu_{1f_{2}} \mu_{f_{2}2}$, respectively, connecting the initial and final states with each other. If these two products have opposite signs, the oscillations of two pairs of absorption lines will be completely out of phase with respect to each other. Indeed, according to our estimations of the relevant dipoles (see Appendix~\ref{appdx:DM}), the relative sign between the dipoles in the region $R_1$ is flipped compared to the $R_2$ lines. For transition $\nu_1$ and $\nu_2$, for $m=1/2$, the dipoles have ``$-$'' and ``$+$'' signs, respectively, while the dipoles for both transitions $\nu_3$ and $\nu_4$ have ``$+$'' signs. Accordingly, the oscillations in the region $R_1$ are out of phase compared to those in region $R_2$. This demonstrates that the time-dependent transient absorption spectroscopy is capable of probing not only the electronic coherences created in the system but also signs of the dipole moments connecting the initial and final states with each other.

\begin{figure*}
\includegraphics[width=0.8\linewidth]{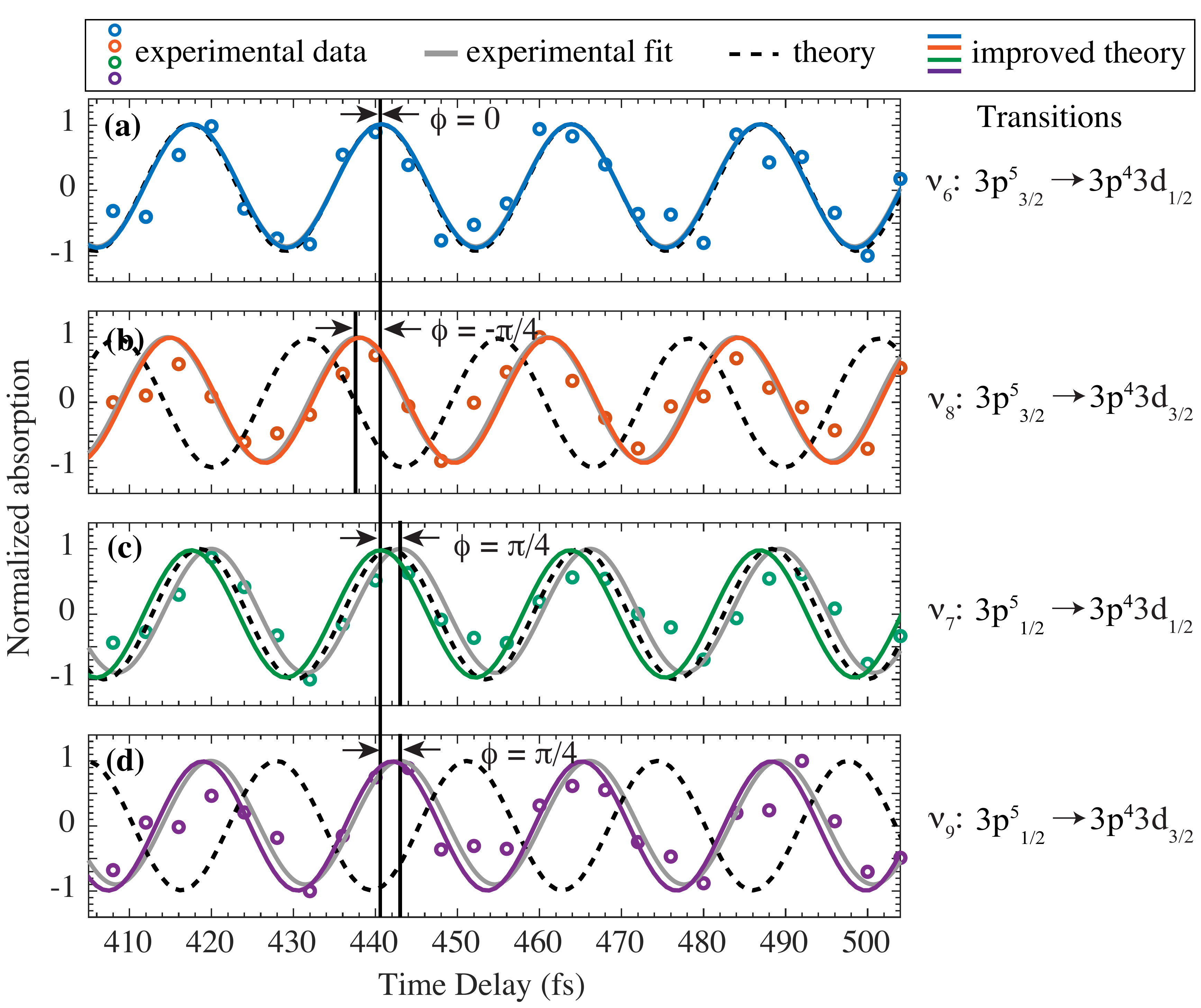}
\caption{Experimentally measured (dots) and theoretically simulated (lines) absorption lineouts for the transitions connecting the spin-orbit split initial $3s^2 3p^5_{J=1/2}$ and $3s^2 3p^5_{J=3/2}$ states to the final $3s^2 3p^4 3d_{J=1/2}$ and $3s^2 3p^4 3d_{J=3/2}$ states. The close proximity of the absorption lines to each other causes phase offsets in the corresponding oscillations. Theoretical simulation results in dashed lines are improved significantly by implementing a change in the sign of the dipole moment associated with the $\nu_8$ transition, capturing the measured oscillation phase offsets, as shown by the solid lines.}
\label{fig:intrachannel}
\end{figure*}

The lineout of $\nu_8$ transition, shown in panel (c) of Fig.~\ref{fig:lineout1}, does not follow the simple logic outlined in the previous paragraph. The phase offset of $\nu_8$ with respect to $\nu_1$ line is found to be about $3\pi/4$ radians. However, this behavior can also be explained by a deeper analysis of Eq.~(\ref{eq:fourlines}). When we applied this expression to analyze the spectral lines in the regions $R_1$ and $R_2$, the energy separation between the lines and the resolution of our experimental setup were sufficient to treat the energy profiles of the measured transitions as distinct. Thus, only two possibilities for the phase relations between the lines existed in this case, namely either being fully in phase or completely out of phase with respect to each other depending on the signs of corresponding transition dipole moments. In contrast, the small energy separation between the pairs of final states contributing to the region $R_3$ leads to overlapping absorption lines. In this case, the gap between the two final states emerges as an additional energy scale which has an impact on the phase of the oscillations. It is seen from Eq.~(\ref{eq:fourlines}) that if the spacing between the energy levels $f_1$ and $f_2$ is smaller than the widths of the corresponding spectral lines, then the lines will overlap and their oscillation phases can differ even if they share the same final state.

The relationship between the phases of the absorption lines in the region $R_3$ forms the next topic of our investigation. Figure~\ref{fig:intrachannel} depicts the lineouts of the four absorption lines $\nu_6$, $\nu_7$, $\nu_8$ and $\nu_9$ present in that region. Experimental data indicated by points in Fig.~\ref{fig:intrachannel} shows that all four lines have clear phase offsets with respect to each other. The results of our  simulations, depicted by dashed lines in Fig.~\ref{fig:intrachannel}, indicate that although the phases indeed vary between the absorption lines, their agreement with the experimental data for lines $\nu_8$ and $\nu_9$ is poor. Interestingly, the deviations in phases between the experimental and theoretical results for these lines are close to $\pi$ radians. Previously, we attributed the complete phase flip between the oscillations in regions $R_1$ and $R_2$  to the difference in the signs of the corresponding dipole moments. This observation prompted us to check the validity of our simple model for computing the transition dipole moments, detailed in Appendix~\ref{appdx:DM}, between the electronic states.

To correct a potential inaccuracy of our approach in obtaining the phases in region $R_3$, we made an \emph{ad hoc} change to the sign of the dipole moment for transition $\nu_8$ in Tab.~\ref{tab:lines}. The lineouts of the absorption lines computed with this ``improved'' version of our computational scheme are depicted by solid lines in Fig.~\ref{fig:intrachannel}. We note that the full spectrogram shown in panel (f) of Fig.~\ref{fig:full_spec} is also plotted using the improved theoretical model. The agreement between the ``improved'' model and the experimental data is remarkable. {In particular the $-\pi/4$ radian phase difference between $\nu_6$ and $\nu_8$, as shown in panels (a) and (b) of Fig.~\ref{fig:intrachannel}, is captured accurately. We attribute this phase difference to the fact that the splitting between states $3p^4 3d_{J=1/2}$ and $3p^4 3d_{J=3/2}$ is small enough for the oscillations in two absorption lines to overlap. The energy dependent phase of each oscillation, impacts the other, leading to a relative shift between the two. Similarly, the $\nu_7$ oscillation phase is impacted by the nearby presence of $\nu_9$. This explains the observed phase difference between the $\nu_6$ and $\nu_7$ oscillations in panels (a) and (c) of Fig.~\ref{fig:intrachannel}, which in principle should have the same phase given that they share a common final state. Furthermore, the same effect explains why $\nu_8$ oscillation is shifted by a fraction of $\pi$ radians relative to $\nu_1$ in Fig.~\ref{fig:lineout1}.} 

\begin{figure*}
\includegraphics[width=\linewidth]{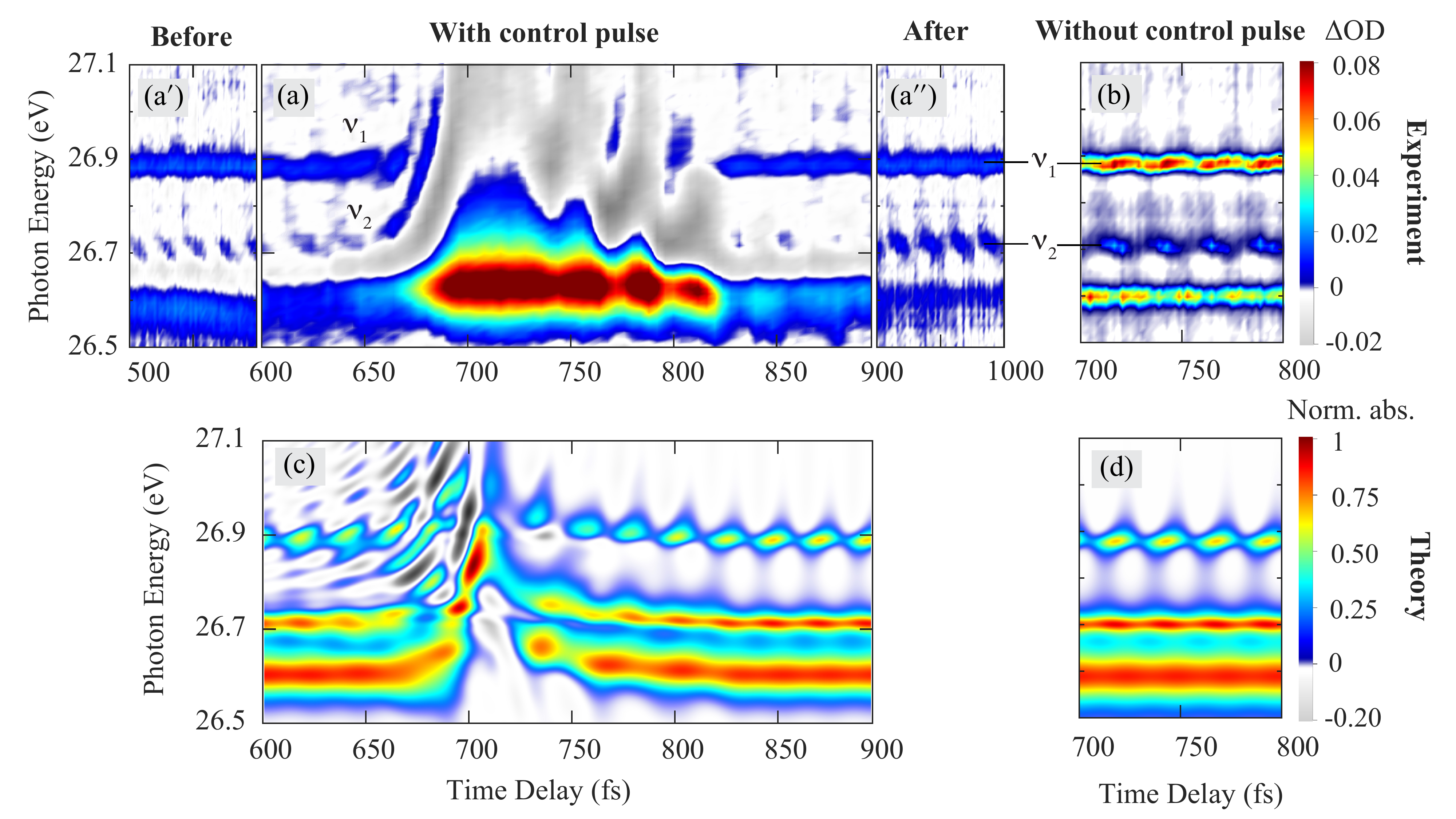}
\caption{Experimentally measured (top panels) and theoretically simulated (bottom panels) absorption spectra of argon ions with (left column) and without (right column) the action of a strong MIR control pulse. The control MIR field leads to the enhancement of neutral absorption and AC Stark shifting of the cationic absorption lines. The electronic coherences created between the spin-orbit split ground states of the cation are transiently impacted by the control pulse leading to the temporal suppression of coherent oscillations over several cycles of the MIR field. After the control pulse, the absorption line oscillations are identical to those measured before the MIR pulse arrival.}
\label{fig:control}
\end{figure*}

Summarizing our observations in this section, we obtained good agreement between the results of our measurements and the simulations. The experimental data guided our theoretical investigations by indicating clearly that for a full correspondence between the experiment and theory some of the dipole moments utilized in the simulations must have an opposite sign. We attribute the discrepancy in our theory to a minimalistic hydrogen-like model used for computing the transition dipole moments. It has been widely regarded that electron correlation effects, which are ignored in the computational approach, can play a substantial role in the  properties of many-electron systems. {While numerous high-level approaches capable of accessing the excited electronic states of atoms have been invented in the past (see, e.g., Ref.~\cite{jonsson2013})}, they are still very limited in their abilities to reach high-lying electronic states such as those measured in our experiment. Thus, our experimental measurements demonstrate the power of the time-dependent transient absorption spectroscopy to highlight the nuanced features of the system under study, beyond straightforward observations of electronic coherences reported before.

\subsection{Control pulse}
\label{sec:results_control}
Next we consider the effect of a third MIR control pulse on the system and discuss the changes in the absorption spectra in the region  $R_1$, accessed by the 21st harmonic of our XUV pulse. 
%In order to generate the control pulse it was necessary to lower the intensity of the ionizing laser. According to our estimations with a simple Keldysh model~\cite{Keldysh1965}, this has the effect of reducing the number of the ionized atoms in half which, in turn, substantially lowers the absorption cross-section for all the measured ionic lines. In addition to this, there are two compounding effects that weaken the sensitivity to other ionic processes. On the one hand, according to data from NIST~\cite{NIST_ASD_2024}, the static photoabsorption cross section for the lines in regions $R_2$ and $R_3$ are substantially weaker than those in region $R_1$. On the other hand, the foil filter used in our experiment reduces the intensity of the 17$^{\text{th}}$ and 19$^{\text{th}}$ harmonics to one ninth and one quarter of the intensity of the 21$^{\text{st}}$ harmonic, respectively. All these factors have cumulative effect leading to a poor signal to noise ratio in the measurements of effects of the control MIR pulse in the regions $R_2$ and $R_3$ of the spectra, while allowing for a clearer signal in the region $R_1$. Therefore, the remainder of this section is devoted to the analysis of the absorption lines belonging to the region $R_1$.
The experimentally measured ATAS spectrum recorded before, during, and after the interaction of the system with the MIR control pulse is shown in panels (a$'$), (a), and (a$''$), respectively, of Fig.~\ref{fig:control}. For comparison, we also show in panel (b) of Fig.~\ref{fig:control} the spectrum obtained in absence of the control field. Results of our simulations, performed using the formalism discussed in Sec.~\ref{sec:theory}, for the system with and without control field are depicted in panels (c) and (d), respectively, of Fig.~\ref{fig:control}.

When the system is exposed to the strong MIR control field we observe a large change in the intensity of one of the absorption lines for the XUV delays between 650 and 850~fs. This effect arises due to the existence of the $3s 3p^6 4p$ autoionizing electronic state around $26.6$~eV. %This state is observable due to geometrical setting of the laser arrangement and the way the change in the optical density is calculated. As was discussed earlier in this work, the reference from which the optical density is quantified is the response of the neutral argon to the XUV pulse, including a strong signal arising from the absorption to this autoionizing state. After the MIR pulse ionizes the system, one would expect the signal to be absent. Nonetheless, since the waist of the MIR ionizing pulse is narrower than the waist of the XUV pulse, the latter pulse will then interact with some residual neutral atoms. This means that mixed with the ionic response to the control pulse the measurements also include the response of the autoionizing electronic states to the strong control pulse.
In our three-pulse experimental setup, the intensity of the ionizing MIR is slightly lower, and this leads to a smaller ionization fraction in the focal volume. As a result, the XUV pulse interacts with some neutrals and populates this autoionizing state.
%\textcolor{blue}{The reason why neutral states are observed at all, is that the focusing region of the ionizing MIR pulse is much tighter than that of the XUV probing pulse. Thus, while the ionization rate is one hundred percent in where the MIR is focused, there are neutral atoms that still interact with the probe.} 
We observed that both the position and the Fano line-shape of the autoionizing line is affected by the control field, which is reminiscent of the strong-field transient absorption. The quantum dynamics of this specific autoionizing state have been explored theoretically and experimentally in Refs.~\cite{Yanez2022,Harkema2021,Cariker2024}. 

%In the current experiment, we observe the shift due to the splitting of the dressed states, owing to the coupling between $3s 3p^6 5s$ and $3s 3p^6 3d$ configurations. A differentiating aspect in the current study is the observation of apparent oscillations in the absorption spectrum. We observe a modulation of the autoionizing line which is distinct from the coherent hole beating observed earlier in the ionic sample, and represents an artifact of the temporal pulse structure.

The control pulse also produces significant changes in the positions of the ionic absorption lines compared to the unperturbed case. During the control pulse, the lines $\nu_1$ and $\nu_2$, at $26.7$~eV and $26.9$~eV, respectively, are transiently shifted upwards, as seen in panel (a).  
%In contrast to the enchantment of the intensity of one of the lines that we attributed to the interaction of the control field with the neutral system, the changes in the energy positions of the lines are due to the response of the ionic system. While the ionic signal is intertwined with the response of the neutral, their contributions add to each other incoherently making it easy to separate these two effects and model the behavior of the ionic system in the presence of the MIR field. 
Panel (c) of Fig.~\ref{fig:control} shows the theoretical calculation of the response of ionic transitions to the control pulse. The upward shift in the energies of the absorption lines is clear, and the physical mechanism behind this lies in the well-known AC Stark effect~\cite{Autler1955,Wu2016,Santra2011}. 
%When the XUV and MIR pulses overlap, the MIR-dressed  states of the system show up at different energies in the spectra, as it has already been demonstrated earlier in Refs.~\cite{Wu2016,Santra2011}. The positions of these dressed states are dependent on the strength of the MIR pulse and thus these lines shift as the XUV excitation takes place at different positions within the action of the control field.

%A more intuitive interpretation of the AC Stark effect arises from the idea that the measured radiation comes from an oscillating dipole formed by the transfer of population from the ground state to the excited states. As the dipole oscillates, it emits radiation with frequency corresponding to the energy difference between the ground and the excited state. As the energy of the excited state continuously shifts due to the strong IR field, the dipole oscillation shifts and the energy range of radiation from the dipole broadens to account for the shift. We can estimate how much it broadens by calculating the expected AC Stark shifts. The lines overlaid on the heatmaps in panels (a) and (c) of Fig.~\ref{fig:control} show how both experiment and theory follow this trajectory. We estimate the Stark shifts by calculating the shift due to the average of the IR field through the duration of the XUV pulse.%, which is just an approximation to the AC stark shift.

After the control pulse, the Stark shifts subside and the coherent oscillation reappears with its relative phase and periodicity unchanged. Importantly, we would like to point out that the oscillations of the ionic absorption lines due to the coherent electronic wave packet created by the SFI process remains nearly unaffected by the action of the control MIR pulse. As one can see from panels (a$'$) and (a$''$) of Fig.~\ref{fig:control}, the oscillations of $\nu_1$ and $\nu_2$ lines are symmetric around the control pulse, suggesting that the effects of the control laser are only transient and do not affect the initial coherent superposition of the ionic ground states. This demonstrates a surprising robustness of quantum electronic coherences to the action of intense external perturbations.

\section{Conclusion}
\label{sec:concl}
In conclusion, our study showcases the experimental capability to generate and measure the ultrafast dynamics in a coherent superposition of the spin-orbit split $3s^2 3p^5_{J=1/2}$ and $3s^2 3p^5_{J=3/2}$ ground electronic state of the argon ion. We observed clear signatures of the created electronic coherences by measuring the time-resolved absorption spectra over a wide range of energies covered by multiple XUV harmonics. The time scale of the measured oscillations has been attributed to the energy separation between the initially populated states. We characterized these oscillations in nine different transitions between the initial and electronically excited final states of the cation. We analyzed the phase shifts between the oscillations of the lines in three regions, which gave us access to detailed information about the dynamics and properties of the system. 

The first region involving the transitions to the $3s^2 3p^4 5d_{J=1/2}$ state, shows two absorption lines oscillating in phase. This scenario, involving two initial and one final state, represents a minimal model where the absorption spectra can show the fingerprints of quantum coherences created between the initial states. 
%Indeed, it has been pointed out earlier~\cite{Santra2011} that the time-resolved absorption spectroscopy is capable of capturing the dynamics in a system only if the initial and final states are appropriately coupled with each other. Although this condition is, in general, not that uncommon, even in atomic systems some of the absorption lines can be useless for observing the created electronic coherences due to the absence of the required transitions.
The second region contains transitions to two spin-orbit split $3s^2 3p^4 4d_{J=3/2}$ and $3s^2 3p^4 4d_{J=5/2}$ states. According to the dipole selection rules both initial states can be coupled to a final state with $J=3/2$, only one of the initial states is coupled to a final state with $J=5/2$. Thus, three absorption lines were observed in the second region. While $J=3/2$ to $J=5/2$ line was expected to be stationary due to the absence of the second interfering path, nevertheless a close overlap with a neighboring $J=3/2$ to $J=3/2$ transition leads to oscillatory behavior. A more striking observation was that the absorption lines in the second region oscillate completely out of phase with respect to the lines in the first region. This previously unobserved behavior is explained by the flip in the signs of one of the dipole moments connecting the initial and final states.

In the third region, we observed even more complicated relations between the phases of four absorption lines connecting the initial states to $3s^2 3p^4 3d_{J=1/2}$ and $3s^2 3p^4 3d_{J=3/2}$ final states. The oscillating absorption lines in this region have considerable overlaps with each other. The observed phase offsets are attributed to the intermingling of neighboring oscillations due to the long tails of Lorentzian-shape absorption profile of each line.

We also studied the response of the system to a strong MIR pulse. We found that part of the experimental absorption includes features coming from the autoionizing $3s 3p^6 4p (^1P)$ state of neutral argon atom. Nevertheless, we were able to disentangle contributions coming from the ionic system. We observed the shifting of the absorption lines due to the AC Stark effect. Importantly, this strong-field control did not influence the coherent superposition of the ionic states.

All these conclusions were supported by a detailed theoretical analysis using analytic expressions derived within the perturbation theory framework. Furthermore, we reproduced all the experimentally measured features in the spectra by performing numerical solutions of the TDSE. Our analysis of the the spectra measured in the third region, demonstrated the intertwining of several effects arising from the dipole matrix couplings, populations of the initial states, and excited state lifetimes. This yields oscillations whose phases vary in a wide range of values depending on the input conditions. Furthermore, these phases can change dynamically with the photon frequency and can superimpose on different lines. We were able to find system parameters that reproduce the experimentally measured oscillations in the spectra, 
%However, a more detailed coherence information, for example, in the form of a two dimensional spectroscopy experiment, would allow us to determine some of this parameters from a different measurement and so identify the appropriate characteristics of the process to model. 
however, theoretical assignment of a physical meaning to the phases of oscillations would also require a more comprehensive treatment of the atom encompassing the ionization step, as well as the complexity of the interaction between the electron-double ionized ion and the complex electronic configurations of the ionic bound states. We leave this for a future study.
%Thus, more deep interpretation of these phase differences lies beyond the scope of this paper, and is left for future experimental and theoretical studies.

Our experimental methods and the supporting theoretical model serve as a step towards for the study of coherent charge migration and electronic wave packet motion control in more complex systems such as molecules. In a paper~\cite{Alarcon2025} published in parallel with this experimental work, we propose an approach to control the coherent superposition of ionic states via stimulated Raman adiabatic passage (STIRAP) technique. Notably, the experimental set up required to perform the proposed measurements is similar to the one used in this study. Our experimental and theoretical work highlights power of the attosecond transient absorption spectroscopy for investigating ultrafast quantum dynamics in matter.

\section{Acknowledgments}

NC, MM, and AS acknowledge support of the U.S. Department of Energy, Office of Science, Office of Basic Energy Sciences, Award No. DE-SC0018251. IS, JW, and DB received support from NSF PHY award 2207641. M.A. and N.V.G. acknowledge support from the U.S. Department of Energy, Office of Science, Office of Basic Energy Sciences, Award No. DE-SC0024182.

\appendix
\begin{widetext}

\section{Evaluation of dipole matrix elements between spin-orbit coupled electronic states of argon cation}
\label{appdx:DM}
To access the excited electronic states of singly ionized argon, we employed a semi-empirical methodology relying on data from the NIST Atomic Spectra Database~\cite{NIST_ASD_2024}. While it is, in principle, possible to compute bound electronic states of atomic systems using high-level \textit{ab initio} methods, the accuracy of these simulations for highly-excited electronic states is questionable and thus the usage of first-principle techniques does not provide clear advantages over a simple semi-empirical approach we describe below.

We import from NIST the energies and the configurations of electronic states. In singly ionized argon, states are classified using both $LS$ and $JK$ angular momentum coupling schemes. A typical electronic configuration of an $LS$-coupled state is specified as
\begin{equation}
    3s^2 3p^4 (^{2S_p+1}L_{p} ) n l
    \quad
    {}^{2S+1}L_J,
\end{equation}
where $3\text{s}^2 3\text{p}^4$ indicates the electronic configuration of the parent doubly charged ion forming a term with $L_p$ and $S_p$ orbital and spin quantum numbers, respectively. The active outer electron is characterized by the principal quantum number $n$ and an orbital angular momentum $l$. The total electronic term is defined by quantum numbers $L$ and $S$. $L$ is formed by coupling $L_p$ with $l$. $S$ if formed by coupling $S_p$ with $1/2$ spin of the outer electron. Finally, $L$ and $S$ are coupled to form a total angular momentum $J$. The projections of $J$ on the $z$ axis are denoted by $M$ in the follow-up derivations.

Similarly, a typical $JK$-coupled state is specified as
\begin{equation}
    3s^2 3p^4 (^{2S_p+1}L_{p{J_p}} ) n l \quad {}^{2}[K]_J,
\end{equation}
where, in addition to the $L_p$ and $S_p$ quantum numbers discussed before, the parent ion is characterized by a total angular momentum $J_p$. The quantum number $K$ is formed by coupling $J_p$ and the orbital angular momentum $l$ of the outer electron. The multiplicity of $JK$ terms for argon cation is always $2$ indicating that only a single outer electron is considered. Finally, this $1/2$ spin quantum number of the outer electron is coupled to $K$ forming the total angular momentum $J$.

%In order to compute the dipole matrix elements, the states we will use are of the form
%
%\begin{equation}
%    \left| \psi_{nlJM} (r,\theta,\phi) \right\rangle = R_{nl}(r) \Phi_{L_pS_p}(\omega) \left|\{...\}J M\right \rangle, 
%\end{equation}
%
%where the curly brackets and ellipses represent either one of the two coupling schemes, $R_{nl}(r)$ is the radial part of the outermost electron, and $\Phi(\omega)$ represents the wave function of the core electrons.\\
%
%Assuming that the described coupling schemes accurately represent the dominant configurations of the corresponding electronic states, we compute the transition dipole moments as following. First of all, we require that the term of the parent ion remains unchanged during the electronic transition thus focusing on the properties of the outermost electron only and assuming zero dipole moment if the parent ion has been modified. Then, the outer electron is characterized by a wavefunction
%
%where $R_{nl}(r)$ and $Y_J^M (\theta,\phi)$ denote the radial and angular parts of the wavefunction, respectively, thus forming the representation of $|n l J M\rangle$ state in spherical coordinates $r$, $\theta$, and $\phi$.

In order to compute the transition dipole moments between the electronic states, we first introduce the notation $|\gamma J M \rangle$, where $\gamma$ denote a set of quantum numbers specific for the coupling scheme under consideration. Expressing the dipole operator along $z$ coordinate using spherical harmonics, such that $\hat{z} = r Y_{1}^0$, we compute the transition between the initial $|\gamma J M\rangle$ and final $|\gamma' J' M'\rangle$ states using the Wigner--Eckart theorem:
\begin{equation}
    \bra{\gamma' J' M'} r Y_{1}^0 \ket{\gamma J M} = (-1)^{J'-M'}
    \begin{pmatrix}
        J'  & 1 & J \\
        -M' & 0 & M
    \end{pmatrix}
    \redmat{\gamma' J'}{r Y_{1}^0}{\gamma J},
\end{equation}
with the reduced matrix element $\redmat{\gamma' J'}{r Y_{1}^0}{\gamma J}$. This reduced matrix element depends on the type of angular momentum coupling utilized for describing the two states of interest. One typically deals with this coupling using standard formulas, like Edmond's 7.1.8~\cite{edmonds1996angular}, but since in our case different states have different angular momentum couplings we simplify the calculations by considering only the hydrogenic term:
\begin{equation}
     \redmat{\gamma' J'}{r Y_{1}^0}{\gamma J}
    \approx \delta_{L_p L'_{p}} \delta_{S_p S'_{p}}(-1)^{l'}\sqrt{2l+1}\sqrt{2l'+1}\sqrt{\frac{3}{4 \pi} }
    \begin{pmatrix}
        l' & 1 & l \\
        0  & 0 & 0 
    \end{pmatrix}
    \int_0^{\infty} R_{n'l'}(r) r R_{nl}(r) r^2 dr,
\end{equation}
thus utilizing only those quantum numbers which are common for both electronic states. The required radial wavefunctions are approximated by hydrogen-like orbitals with appropriate quantum numbers and nuclear charge $Z=2$, indicating that we are considering a single outer electron interacting with the doubly charged parent ion.

\section{Time-dependent transient absorption cross-section with a Gaussian pulse}
\label{appdx:gauss}
Following the discussion presented in Ref.~\cite{Golubev2021}, here we derive a perturbative expression for the absorption cross-section for the case when the probe pulse has a Gaussian profile:
\begin{equation}
\label{eq:Gaussian}
    E(t) = E_0 e^{-\left(\frac{t-\tau}{\gamma}\right)^2} \cos [\omega_0(t-\tau)],
\end{equation}
where $E_0$ is the magnitude of the field, $\tau$ denotes the center of the Gaussian envelope, $\gamma$ is the parameter controlling its width, $\omega_0$ is the photon energy, and the CEP of the field is neglected for simplicity. The Fourier transform of Eq.~(\ref{eq:Gaussian}) is given by
\begin{equation}
    \tilde{E}(\omega) = E_0 \frac{\gamma}{\sqrt{8}} e^{i \omega \tau} \left[e^{-\frac{\gamma^2}{4}(\omega-\omega_0)^2}+e^{-\frac{\gamma^2}{4}(\omega+\omega_0)^2}\right].
\end{equation}

To obtain the absorption cross-section, we use a familiar expression
\begin{equation}
    \sigma(\omega) = \frac{4 \pi \omega}{c} \text{Im} \frac{\tilde{P}(\omega)\tilde{E}^*(\omega)}{|\tilde{E}^*(\omega)|^2},
\end{equation}
where $\tilde{P}(\omega)$ is the Fourier transform of the polarization response
\begin{equation}
    P(t) = \bra{\psi(t)}\hat{\mu}\ket{\psi(t)},
\end{equation}
computed with respect to the time-dependent quantum state $\ket{\psi (t)}$. Assuming that the spectrum of the atomic Hamiltonian $\hat{H}_0$ is known, such that $\hat{H}_0 |\psi_k\rangle = \epsilon_k |\psi_k\rangle$, the time-dependent quantum state in the field-free regime can be expanded as
\begin{equation}
    |\psi(t)\rangle = \sum_k c_k e^{-i\epsilon_k t } |\psi_k\rangle.
\end{equation}

Employing the first-order time-dependent perturbation theory to approximate the evolution of the system under the action of the Hamiltonian $\hat{H}(t)=\hat{H}_{0}-\hat{\vec{\mu}} \cdot \vec{E}(t)$, one can express the first order polarization correction as
\begin{equation}
    P^{(1)}(t) = -i \sum_{k,j,f} c_k^* c_j 
        \bra{\psi_k}\hat{\mu}\ket{\psi_f} \bra{\psi_f}\hat{\mu}\ket{\psi_j} 
        \int_{-\infty}^{t} e^{i(\epsilon_k-\epsilon_f)t+i(\epsilon_f-\epsilon_k)t'}E(t') dt'+\text{c.c.},
\end{equation}
where we utilized the resolution of identity $\sum_f \ket{\psi_f} \bra{\psi_f} \equiv 1$ to simplify the expression and $\text{c.c.}$ denotes the complex conjugation. Using the Fourier convolution theorem and expanding energies of the final states to a complex domain, $\epsilon_f \to \tilde{\epsilon}_f$, one can show that the Fourier transform of the polarization is given by
\begin{equation}
\begin{aligned}
    \tilde{P}^{(1)}(\omega) =& \frac{E_0 \gamma}{\sqrt{8}} \sum_{k,j,f} c_k^* c_j e^{i\tau(\epsilon_j-\epsilon_k)} 
        \bra{\psi_k}\hat{\mu}\ket{\psi_f} \bra{\psi_f}\hat{\mu}\ket{\psi_j} \\
    \times & \left[
        \frac{
            e^{-\frac{\gamma^2}{4}(\omega-\epsilon_k+\epsilon_j-\omega_0)^2}+e^{-\frac{\gamma^2}{4}(\omega-\epsilon_k+\epsilon_j+\omega_0)^2}
        }{
            \tilde{\epsilon}_f-\epsilon_k-\omega
        } e^{i\omega\tau}
    \right.\\[5mm]
    + & \left.
        \frac{
            e^{-\frac{\gamma^2}{4}(-\omega-\epsilon_j+\epsilon_k-\omega_0)^2}+e^{-\frac{\gamma^2}{4}(-\omega-\epsilon_j+\epsilon_k+\omega_0)^2}
        }{
            \tilde{\epsilon}_f^*-\epsilon_j+\omega
        } e^{i\omega\tau}
    \right].
\end{aligned}
\end{equation}
Under the assumption that only the terms involving $\omega-\omega_0$ in the exponent are relevant, the expression for the absorption cross-section is given by:
\begin{equation}
\begin{aligned}
    \sigma(\omega,\tau) = & \frac{4\pi \omega}{c} \text{Im}\sum_{k,j,f} c_k^* c_j e^{i\tau(\epsilon_j-\epsilon_k)} 
        \bra{\psi_k}\hat{\mu}\ket{\psi_f} \bra{\psi_f}\hat{\mu}\ket{\psi_j} \\
    & \times e^{-\frac{\gamma^2}{4}(\epsilon_k-\epsilon_j)^2}
    \left[
        \frac{
            e^{-\frac{\gamma^2}{2}(\omega-\omega_0)(\epsilon_j-\epsilon_k)}
        }{
            \tilde{\epsilon}_f-\epsilon_k-\omega
        }
        +\frac{
            e^{-\frac{\gamma^2}{2}(\omega-\omega_0)(\epsilon_j-\epsilon_k)}
        }{
            \tilde{\epsilon}_f^*-\epsilon_j+\omega
        }
    \right].
\end{aligned}
\end{equation}
The obtained expression, given by Eq.~(\ref{eq:TA}) in the main text, serves as the starting point for deriving various limiting cases that we use to explain the features observed in the experimentally measured absorption spectra. To simplify the discussion, in the main text we further assumed the uniform energy profile of the probe pulse thus switching from the Gaussian to a delta pulse. For completeness, here we present expressions analogous to those derived in Sec.~\ref{sec:results_two_pulse} but assuming the probe pulse has a Gaussian profile.

To analyze the dependence of the phase of the absorption signal on the photon energy $\omega$, we express each of the absorption lines as 
\begin{equation}
    \frac{1}{\tilde{\epsilon}_f - \epsilon_k-\omega} = L_{fk}(\omega)e^{i\phi_{fk}(\omega)},
\end{equation}
where the real quantities $L_{fk}(\omega)$ and $\phi_{fk}(\omega)$ are given by 
\begin{equation}
    L_{fk}(\omega) = \sqrt{\frac{1}{(\epsilon_f-\epsilon_k-\omega)^2+\Gamma_f^2/4}} \quad \text{and}\quad \phi_{fk}(\omega)=\arctan \frac{\Gamma_f/2}{\epsilon_f-\epsilon_k-\omega},
\end{equation}
and $\epsilon_f$ and $\Gamma_f$ denote the real and imaginary parts of final state energies $\tilde{\epsilon}_f=\epsilon_f-i\Gamma_f /2 $. The simplified formula for the absorption cross-section can now be written as
\begin{equation}
\begin{aligned}
    \sigma(\omega,\tau) =\frac{4\pi \omega}{c} \text{Im} 
        & \sum_{k,j,f} c_k^* c_j e^{i\tau(\epsilon_j-\epsilon_k)} \bra{\psi_k}\hat{\mu}\ket{\psi_f} \bra{\psi_f}\hat{\mu}\ket{\psi_j} \\
        \times & e^{-\frac{\gamma^2}{4}(\epsilon_k-\epsilon_j)^2} e^{-\frac{\gamma^2}{2}(\omega-\omega_0)(\epsilon_j-\epsilon_k)}
    \left(
        L_{fk}(\omega)e^{i\phi_{fk}(\omega)}
        +L_{fj}(-\omega)e^{i\phi_{fj}(-\omega)}
    \right).
\end{aligned}
\end{equation}

Considering the case where there are two initial states, $i,j \in {1,2}$, and only one final state $f$, the oscillating portion of the absorption cross-section is given by
\begin{equation}\label{eq:ap1}
\begin{aligned}
    \sigma_{\text{osc}}(\omega,\tau) = \frac{4\pi \omega}{c} |c_1||c_2| e^{-\frac{\gamma^2}{4}\epsilon_{12}^2} \mu_{f1} \mu_{f2}
        \Bigl\{
              & e^{-\frac{\gamma^2}{2}(\omega-\omega_0)\epsilon_{12}}L_{f1}(\omega) \sin\left[\epsilon_{12}\tau-\tau_{12}+\phi_{f1}(\omega)\right] \\
            - & e^{+\frac{\gamma^2}{2}(\omega-\omega_0)\epsilon_{12}}L_{f2}(\omega)  \sin\left[\epsilon_{12}\tau-\tau_{12}-\phi_{f2}(\omega)\right]
        \Bigl\},
\end{aligned}
\end{equation}
where we denoted $c_1^* c_2 = |c_1||c_2|e^{i\tau_{12}}$, $\epsilon_1-\epsilon_2=\epsilon_{12}$, and $\bra{\psi_k}\hat{\mu}\ket{\psi_f} = \mu_{kf}$.

Including an additional final state, thus considering transitions between the two initial states $i,j \in {1,2}$ and two final states $f_1,f_2$, yields
\begin{equation}\label{eq:ap2}
\begin{aligned}
    \sigma_{\text{osc}}(\omega,\tau) = \frac{4\pi \omega}{c} |c_1||c_2| e^{-\frac{\gamma^2}{4}\epsilon_{12}^2} 
    \Bigl\{
          & e^{-\frac{\gamma^2}{2}(\omega-\omega_0)\epsilon_{12}}L_{f_11}(\omega) \mu_{f_1 1}\mu_{f_1 2}\sin\left[\epsilon_{12}\tau-\tau_{12}+\phi_{f_11}(\omega)\right] \\
        + & e^{-\frac{\gamma^2}{2}(\omega-\omega_0)\epsilon_{12}}L_{f_21}(\omega) \mu_{f_2 1} \mu_{f_2 2} \sin\left[\epsilon_{12}\tau-\tau_{12}+\phi_{f_21}(\omega)\right] \\
        - & e^{+\frac{\gamma^2}{2}(\omega-\omega_0)\epsilon_{12}}L_{f_12}(\omega) \mu_{f_1 1} \mu_{f_1 2}\sin\left[\epsilon_{12}\tau-\tau_{12}-\phi_{f_12}(\omega)\right] \\
        - & e^{+\frac{\gamma^2}{2}(\omega-\omega_0)\epsilon_{12}}L_{f_22}(\omega)\mu_{f_2 1} \mu_{f_2 2}\sin\left[\epsilon_{12}\tau-\tau_{12}-\phi_{f_22}(\omega)\right]
    \Bigl\}.
\end{aligned}
\end{equation}
\end{widetext}

\bibliography{refsAr_ions}

\end{document}